# Unveiling Defect Physics in Gapped Metals: A Theoretical Investigation into Defect Formation and Electronic Structure Interplay

Harshan Reddy Gopidi[1], Lovelesh Vashist[1], and Oleksandr I. Malyi[1,#]

[1]Centre of Excellence ENSEMBLE[3] Sp. z o. o., Wolczynska Str. 133, 01-919, Warsaw, Poland

[#]**Email:** oleksandrmalyi@gmail.com

**Abstract:**
In materials science, point defects play a crucial role in materials properties. This is particularly well known for the wide band gap insulators where the defect formation/compensation determines the equilibrium Fermi level and generally the doping response of a given material. Similarly, the main defect trends are also widely understood for regular metals (e.g., Cu and Zn discussed herein). With the development of electronic structure theory, a unique class of quantum materials – gapped metals (e.g., $Ca_6Al_7O_{16}$, $SrNbO_3$, $In_{15}SnO_{24}$, and $CaN_2$) that exhibit characteristics of both metals and insulators - has been identified. While these materials have internal band gaps similar to insulators, their Fermi level is within one of the main band edges, giving a large intrinsic free carrier concentration. Such unique electronic structures give rise to unique defect physics, where the formation of acceptor or donor defect directly affects not only the electronic structure but also can substantially shift the Fermi level. Motivated by this, herein, we develop a fundamental first-principles theory of defect formation in gapped metal with a primary focus on accurate calculations of defect formation energy. We demonstrate that due to electron-hole recombination, the formation of acceptor defects in n-type gapped metals results in significant dependence of defect formation energy on supercell size, which we explain by the change of band filling and its effect on the defect formation energetics. To accurately describe defect formation energy, we revisit the phenomenology of band-filling corrections and demonstrate the effect of this correction, accurate potential alignment, and other factors that can affect defect energetics. Thus, this work not only sheds light on the intrinsic properties of gapped metals but, in general, establishes a theoretical foundation for analyzing defects in gapped metals.

***Defects as a cornerstone of materials properties:*** Every material contains point defects occurring as the result of an interplay between the energy cost to introduce change in local bonding environments and an increase in configuration entropy[1-3]. While point defects can be considered as imperfections of the solids, it turns out that often materials properties are defined by them and their interactions. This is especially well noted for the insulator where defect compensation defines the position of the equilibrium Fermi level[4-10], the n- and p-type nature of a compound[6, 11] and even the doping response of a solid[4, 6, 9, 12]. For instance, the majority of the oxides (e.g., ZnO) cannot be made p-type due to the low position of the principal valence band maximum[13-15]. This led the research community to use the electronic structure theory as a common tool to investigate defect physics, forming the concept of defect transition levels, n- and p-type pinning energy[11], charge neutrality rule[16], H pinning energy[14-15], or even more generally that formation (concentration) of point defect is defined not only by the defect itself but by other defects via accounting for the charge neutrality rule. Hence, modern experimental methodologies, such as atomic-resolution transmission electron microscopy, scanning tunneling microscopy, positron annihilation spectroscopy, deep-level transient spectroscopy, and electron paramagnetic resonance, are extensively employed to elucidate, detect, and scrutinize defects with high accuracy. These techniques are particularly potent when synergized with density functional theory (DFT) analysis of point defects that often can provide vital details on specific defect properties (e.g., transition levels)[17].

***The need for accounting of post-process corrections in the electronic structure theory of defects:*** Within the defect theory, the main idea of DFT calculations is to use small simulation cells and reproduce properties of defects in the dilute limit. However, it becomes clear that the description of the formation energy of point defects can directly be affected by simulation cells. Thus, periodic boundary conditions can influence the behavior of defects in several significant ways: (i) defect can induce artificial interaction due to defect-induced local strain[18-19]; (ii) supercell with charge defects has strong charge-charge interaction[20-23]; (iii) adding electrons or holes to the system can result in artificial band filling when for instance instead of occupation of the bottom of conduction band the carrier addition result in occupation band within non-negligible energy range[24]. It is worth noting that the inaccuracies brought about by these finite-size effects can have profound consequences for the interpretation of physical results and the understanding of potential practical applications of the proposed materials. We emphasize that all the above size effects come from the small supercell size and, in principle, can be avoided by scaling defect formation energy with supercell size[25-27] or applying specific post-energy corrections[20, 24, 26, 28-30]. Indeed, the power of both approaches has been demonstrated for a number of insulators in understanding experimental results - researchers have been able to unravel the microscopic mechanisms governing defect behavior[31-33], color centers[34-36], and other crucial processes within insulators.[6, 11]

**Common beliefs on electronic structure theory of defects in insulators vs metals:** In traditional insulators, the accurate depiction of defects typically necessitates a comprehensive approach that includes various corrections and the use of an exchange-correlation functional capable of precisely replicating band gap energy, as cited in relevant literature[25-26, 37]. In the case of metal, however, the common beliefs are different: (i) one needs to minimize the strain interaction between defects with sufficiently large supercell size; (ii) there are no charged defects in metals as in contrast to a defect-dependent definition of Fermi level in insulators, Fermi level in metal is directly calculated from a number of free carriers - i.e., adding or removing an electron from the system result not in the formation of charge defect but adding or removing electrons to Fermi level[38]; (iii) band filling correction is explicitly not accounted because of continuous nature of bands in metals. Because of this, it is common that there is no post-process correction used in calculations of defect formation energy in metals as long as the supercell size in each direction is larger than about 10 Å to minimize indirect interaction between the defect and its periodic images.

**Not all metals are alike and strong size-dependent defect formation energy is possible:** In traditional metals, the formation energy of point defects (Fig. 1a) is weakly dependent on supercell size after reaching separation of the point defects around 10 Å, which is indeed in good agreement with our first-principles calculations for Cu (space group (SG): Fm$\bar{3}$m (225)) and Zn (SG: P6$_3$/mmc (194)) systems shown in Fig. 1b. This behavior is not surprising as metals typically have more delocalized electron distributions and can easily accommodate strains due to their malleability as compared to regular insulators (e.g., ZnO). Moreover, the free electrons can screen long-range interactions when, for instance, a charge is introduced to the system. This electronic screening is much weaker in insulators due to the lower dielectric constant. As a result, the defect formation energy in metals tends to be less sensitive to supercell size. We note, however, that herein, we find that for some metals, defect formation energy has strong supercell size dependence. For instance, in Fig. 1b, we demonstrate that for some metals, the change in defect formation with supercell size is significant even when the effective lateral dimension of the system is larger than 10 Å. For instance, for $Ca_6Al_7O_{16}$ (SG: I$\bar{4}$3d (220)), the Al vacancy formation energy (calculated within accounting for any corrections, Fig. 1a,b) for 58 and 464-atom supercells differs by 3.14 eV. Similar behavior is observed for $SrNbO_3$:$V_{Nb}$, $In_{15}SnO_{24}$:$V_{In}$, and $CaN_2$:$V_{Ca}$ (see details on the dataset of used materials in methods).

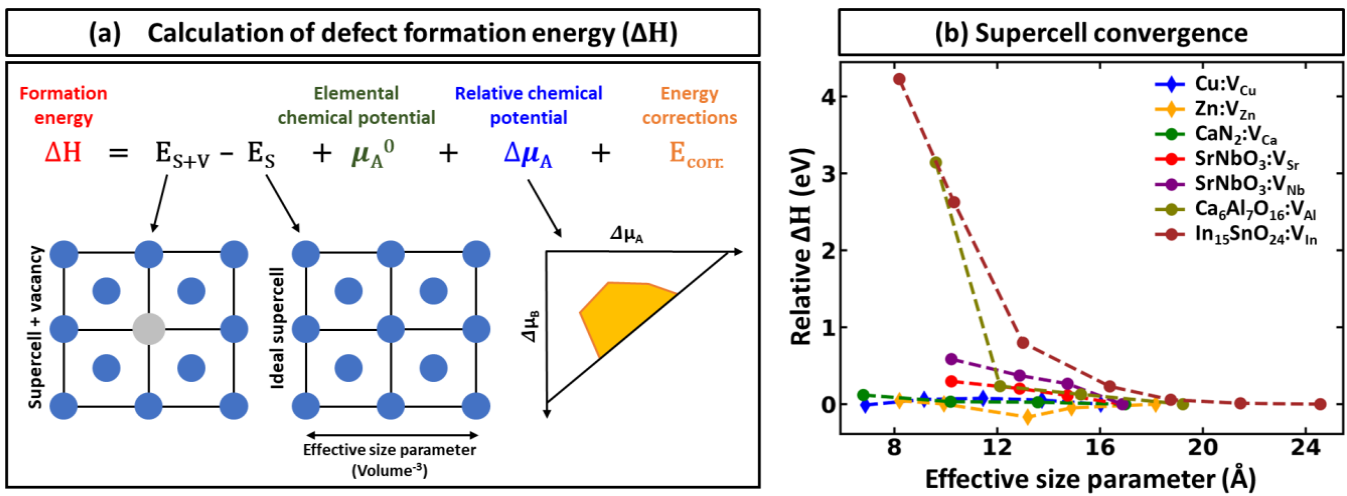

**Figure 1. Calculation of defect formation energy for different metals.**
(a) Schematic for calculations of defect formation energy for metals for different supercell sizes. (b) Relative defect formation energy calculated as the difference of defect formation energy for a given supercell and that for the largest supercell considered for a given system. The figure (a) is inspired by Ref. [29].

**Size dependence of defect formation energy does not originate from structural changes within different supercells or interaction of localized states:** One may naively think that a change in defect formation energy is coming due to a change of internal structure - i.e., depending on supercell size different local structural displacements are stabilized and one needs significantly large supercell size to converge such structural distortion. However, the change of defect formation energy in order of couple eV is significantly larger than that observed for typical defect relaxation[18-19]. To demonstrate this, we analyze atomic displacement in the first coordination sphere for different supercell sizes (Figs. 2a and 2b). The results show that with increasing supercell size, the relative atomic displacements converge. Here, it should be noted that for most of the

systems starting slightly above 10 Å (except the SrNbO$_3$:V$_{Nb}$), all atomic displacements are close to the approximate values. Importantly, a comparison of the results between Fig. 1b and Fig. 2b demonstrates that the convergence of structural properties is reached significantly faster than the convergence of defect formation energy. For instance, increasing the supercell size from 160 to 320 atoms for In$_{15}$SnO$_{24}$:V$_{In}$ results only in a difference in relative relaxation for the first coordination sphere of ~0.006 Å, while the corresponding change in defect formation energy is 0.57 eV. These results thus imply that while the convergence of internal atomic displacements can result in a change of defect formation energy[18], it does not appear to be the driving force for the supercell size dependence of defect formation energy in the considered metals.

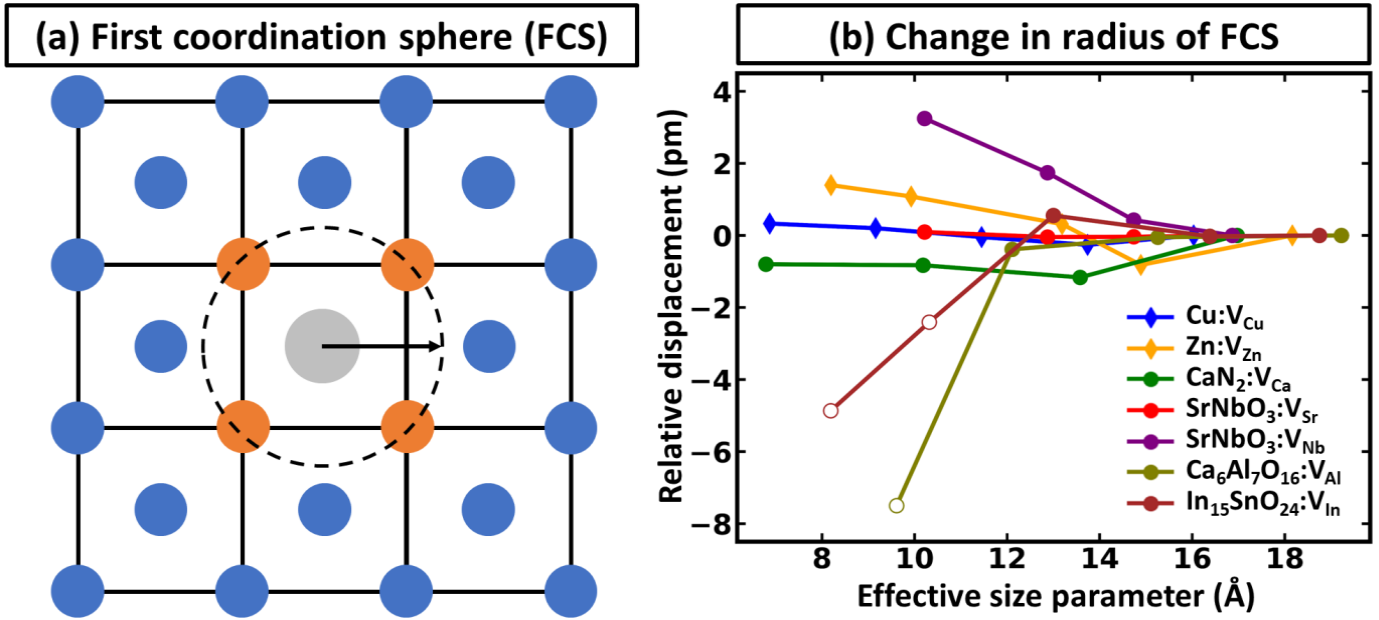

**Figure 2: Defect-induced displacements in metals.**
(a) Schematic illustration of the distribution of local neighbors around the vacancy site showing the first coordination sphere. (b) Relative atomic displacement with respect to supercell size for different metals (hollow circles indicate a system where defect formation removes all free carriers from the principal conduction band. Relative displacement is with respect to the displacement in the largest supercell for each system.

**Gapped metals as a unique class of quantum materials:** To understand better the possible origin for the size dependence of defect formation energy, we refer to the electronic structures (Fig. 3) of the different metallic compounds shown in Fig. 1b. We see that all metals have similar electronic properties with presence of free carriers. What makes the substantial difference, however, is that in Cu and Zn the Fermi level crosses the part of the continuous band (Fig. 3a,b) containing all valence electrons with pseudo-core states located significantly below it. In contrast, SrNbO$_3$, In$_{15}$SnO$_{24}$, Ca$_6$Al$_7$O$_{16}$, and CaN$_2$ also have free carriers, but the Fermi level is not located in the continuous region of the band structure like in Cu and Zn cases. These compounds are known as gapped metals[38-40] – a unique class of quantum materials superposing the existence of a large internal band gap and Fermi level inside of one of the principal band edges. Thus, SrNbO$_3$, In$_{15}$SnO$_{24}$, and CaN$_2$ are n-type gapped metals having Fermi levels inside of the principal conduction band and a large internal band gap below it (Fig. 3c,d,f). At the same time, Ca$_6$Al$_7$O$_{16}$ is an intermediate band gap compound with a partially occupied band (Fig. 3e), for simplicity of further discussion, we will also refer to this compound as n-type gapped metal. We note that all these compounds belong to a wide family of materials that recently attracted significant attention as potential electrides[41-44], thermoelectrics[40, 45], and transparent conductors[38-39, 46-48]. These thus suggest that while gapped metals have metallic properties (e.g., increased resistivity with temperature), they also have inherent properties for insulators (e.g., presence of internal gap between principal band edges). Moreover, in these compounds, only a fraction of valence electrons contributes to free carrier concentration. For instance, $SrNbO_3$ has 1e per formula unit in the principal conduction band with an internal band gap energy of 2.5 eV. Indeed, for all these compounds, the free carrier concentration can be estimated using the sum of composition-weighted common oxidation states (e.g., in the case of $SrNbO_3$, the common oxidation states are +2, +5, and -2 for Sr, Nb, and O, respectively).

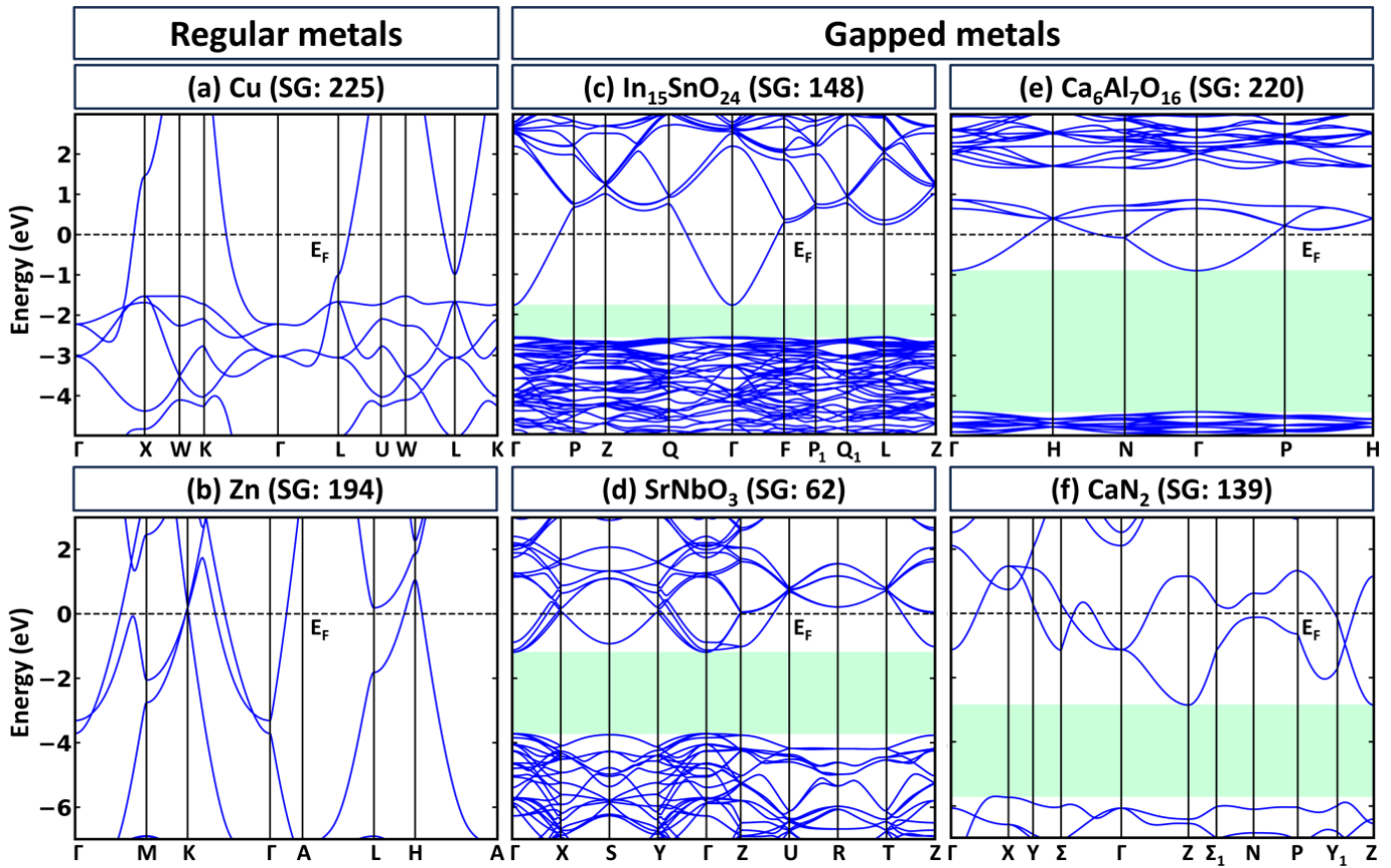

**Figure 3. Electronic properties of different types of metals:**
Electronic band structures of (a,b) regular metals and (c-e) gapped metals as computed using PBE exchange-correlation functional. Space group number is abbreviated as SG. $E_F$ is the Fermi level.

**The electronic structure of gapped metal has unique consequences on defect physics:** In conventional metals such as Cu and Zn, the formation of a vacancy occurs when an atom is removed, thus leaving a "void" in the lattice structure. This void can be occupied by an electron from the "*electron sea*". Moreover, in these metals, all atoms are structurally equivalent. Hence, the generation of a vacancy leads to the removal of both the associated electrons and their corresponding energy states from the electronic configuration, resulting in no significant change in the Fermi level. On the other hand, with gapped metals, vacancy formation, such as that of an acceptor, can lead to pronounced changes in electronic properties. This is because the newly-formed acceptor state introduces unoccupied states below the Fermi level, which can cause some conductive electrons to move from the main conduction band to the acceptor level (a similar mechanism also exists for p-type gapped metals – i.e., compounds having a Fermi level in the principal valence band and a large internal band gap above it[49-50]). This shift can lead to a significant change in carrier concentration, especially in small supercells. For instance, in $SrNbO_3$, a single Nb vacancy removes 5e per vacancy from the principal conduction band. Similar effects are noted in other gapped metals. Specifically, Ca vacancy in $CaN_2$ removes 2e, Al vacancy in $Ca_6Al_7O_{16}$ removes 3e, and In vacancy in $In_{15}SnO_{24}$ removes 3e from the principal conduction band. While these results may be fully visible in the case of high concentration (which indeed can be realized in some gapped metals[38, 50-52]), this behavior implies that an accurate description of defect formation energy at a dilute regime should account for a reduction of free carrier concentration. For instance, in Figure 4b, one can see how the defect formation results in a change of the occupied part of the principal conduction band ($\Delta E_{CB}$) in $Ca_6Al_7O_{16}$. Specifically, the formation of Al vacancy in a 116-atom supercell reduces $\Delta E_{CB}$ to 0.43 eV, while the corresponding $\Delta E_{CB}$ for a 464-atom supercell is 0.89 eV. In the case of certain supercells, the creation of vacancies may lead to a transition from metal to insulator or even a shift of the Fermi level from the principal conduction band to the principal valence band, as exemplified in Fig. 2b and Fig. 4b. This is akin to the behavior seen in insulators where the formation of donor or acceptor defects induces occupation in the conduction or valence band over an artificially wide energy range, differing only in gapped metals the reference state already has Fermi level in one of the principal band edges. This behavior is crucial for understanding the electronic structure and inherently impacts defect energetics, as the energy levels of occupied states are integral to the extension of DFT energy, indicating that each alteration in occupation seen across different supercells will directly influence the defect formation energy. Moreover, for a given compound, a larger shift in the Fermi level due to defect formation results in a larger change of defect formation energy.

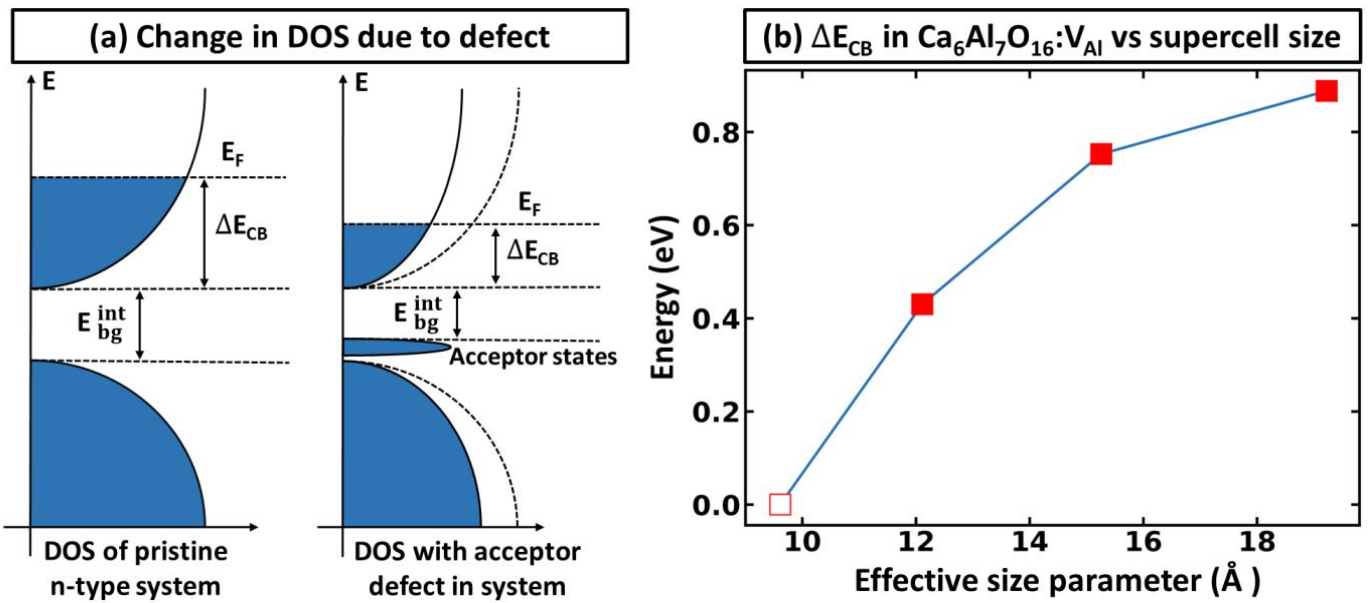

**Figure 4: Effect of acceptor defect formation on electronic properties of gapped metals.**
(a) Schematic illustration of the electronic structure of n-type gapped metal for pristine and acceptor defect systems (where $E_{bg}^{int}$ is the internal band gap, $\Delta E_{CB}$ is the occupied conduction band width, and $E_F$ is the Fermi level). (b) $\Delta E_{CB}$ vs supercell size in $Ca_6Al_7O_{16}:V_{Al}$ system containing single Al vacancy (hollow square is when all electrons are completely removed from principal conduction band).

**Physics of band-filling correction in gapped metals and other factors that can affect calculations of defect formation energy:** One may wonder what the impact of the electronic properties on the supercell-size dependence of defect formation energy is. In traditional insulators, the problem of band-filling is usually solved using either the supercell scaling (it should be noted that scaling is usually done to account for all

factors at the same time and is not specific for band filling) or applying post-process band-filling correction[24, 26]. In the case of the scaling approach, the defect formation energy for the range of supercells is extrapolated to the dilute limit based on the guess of potential physical phenomena controlling the process. In the band-filling correction scheme, the main idea is to restore the expected occupation corresponding to the dilute limit, i.e., in gapped metals, defect formation should not affect the Fermi level in the dilute limit. This expected energy correction thus can be calculated by summation over the eigenvalues as:

$$\Delta_{Band\,filling\,correction} = -\Sigma_{n,k}[\Theta(e_{n,k} - E_F)\omega_k\gamma_{n,k}(e_{n,k} - E_F) + \Theta(E_F - e_{n,k})(\omega_k(1 - \gamma_{n,k})(E_F - e_{n,k})]$$

where $\Theta(x)$ is the Heaviside step function, $\omega_k$ are the weights of the k-points, $e_{n,k}$ are the eigenenergies of state (n, k), $\gamma_{n,k}$ are the occupations of the eigenstate (n, k), $E_F$ is the fermi level of the defect supercell with infinite volume (supercell without defects). We note, however, that, in practical scenarios, the emergence of point defects affects all eigenvalues (Figs. 5a,b). This is because plane-wave first-principles codes do not have common reference states for eigenvalues and hence the band-filling correction formalism indicated above cannot be straightforwardly employed unless a proper alignment of eigenvalues for both defective and pristine systems is made. To illustrate the complexity of such calculations, we consider the formation of Nb and Sr vacancies in the 160-atom $SrNbO_3$ supercell. As shown in Figs. 5a and 5b, both these defects are acceptors but remove different numbers of electrons from the principal conduction band. Specifically, Sr vacancy removes only 2e per vacancy from the conduction band, compared to 5e for the Nb case. The analysis of the electronic density of states shows that the defect formation changes the absolute eigenvalues for both systems compared to those for the pristine case. In our previous work[24], we demonstrated that one could use the electrostatic potentials at the cores of the most remote atoms to align the energy levels of pristine and defective systems and, in this way, efficiently describe the band-filling correction in regular insulators like ZnO. For gapped metals, such alignment is critical as small inaccuracies in its calculations can result in a noticeable change in defect formation energy due to the fact that the gapped metals have high free carrier concentration. Moreover, since different atomic identities are involved (i.e., distinct cations and anions), their electrostatic potentials may be affected differently. For instance, for $SrNbO_3$:Nb, potential alignment using electrostatic potentials for the most remote O vs the most remote Nb atoms results in a difference in the relative defect formation energy of 0.17 eV for 160-atom supercell (Fig. 5c). This trend remains consistent for other supercell sizes, as shown in Fig. 5d. As supercell size increases, the defect formation energies calculated for different methods of potential alignments approach to each other. As noted above, the main cause for such behavior is different shift of electrostatic potential for different atomic identities. Indeed, for some of the considered systems (Fig. 6), there is no substantial difference in the results for different alignment methods. While a more detailed study is needed, we find that this behavior is observed for systems having distinct elemental contributions to band edges - for $SrNbO_3$, the principal conduction band is primarily dominated by Nb-d states, while the corresponding contributions from Sr and O are minimal. This thus results in a fundamental question: what approach for band alignment should one choose for calculating defect formation energy?

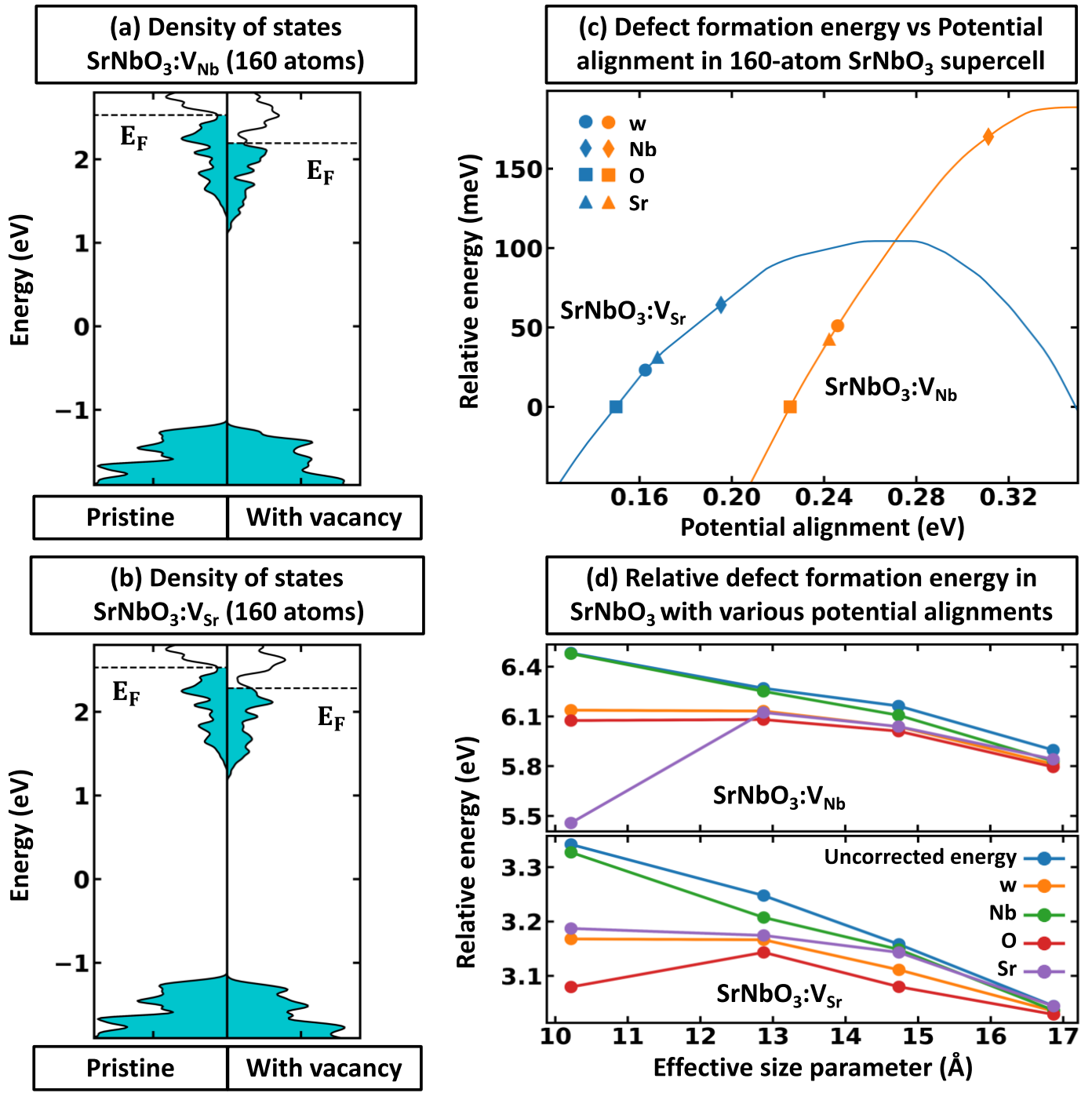

**Figure 5: Strong dependence of defect formation energy on potential alignment.**
Density of states (DOS) for a 160-atom $SrNbO_3$ supercell is presented in two scenarios: (a) with and without Nb vacancy and (b) with and without Sr vacancy. These scenarios illustrate shifts in both the Fermi level and the eigenvalues. (c) Relative band filling correction vs potential alignment, where Sr, Nb, O are averages over 3 most remote atoms of each kind and w is composition weighted average of Sr, Nb, and O potential alignments for 160-atom $SrNbO_3$ supercell containing Sr or Nb vacancy. For each system relative energies are with respect to corresponding O potential alignments. (d) Relative defect formation energy for Sr and Nb vacancy as a function of supercell size using different methods of potential alignment. Relative energy for each system is with respect to an arbitrary energy reference state.

To answer the above question, we recall that the main idea behind band-filling correction is to mimic dilute limit in defect calculations. This practically means that for atoms located far away from the defect, the projected electronic structure should remain unaffected and that for most remote atoms, not only the position of the Fermi level but also the position of band edges should remain unperturbed. Unfortunately, this behavior is often not fully achievable in supercell calculations. Because of this, herein, using a small dataset of gapped metals, we explored different ways for band alignment and identified that composition weighted average electrostatic potential for the most remote atoms from the defect provides the most consistent defect formation trends among different systems (Fig. 6a, b). The composition-weighted alignment can be motivated in the following way: change in the electrostatic potential of different atoms is different, and this results in the optimal change in the electrostatic potential for the entire system to be the composition-weighted average of the change in electrostatic potential of different atoms.

We note, however, that while it is crucial to incorporate band-filling corrections for precise defect formation energy calculations in gapped metals, it is important to acknowledge other factors that may affect defect energetics. Thus, even in traditional metals, the defect formation energy is susceptible to variations in k-point density. Indeed, for Zn, for some supercells, the error bar is in the order of 0.1 eV as compared to converged defect formation energy sizes with ultra-dense k-point grid (i.e., 10,000 vs 100,000 k-points per reciprocal atom). For gapped metals, considered herein, this is a less critical factor, but for other gapped metals, it may play a more significant role. Furthermore, as noted above, defect formation can result in substantial relaxation (e.g., due to breaking strong anion-cation bonds) that may affect defect formation energy for some systems. For $SrNbO_3$:$V_{Nb}$, the difference in relaxation energy (i.e., computed as the energy difference of unrelaxed and relaxed systems containing the defect) computed for 80 and 360-atom supercells is 0.17 eV. Taking into account that relaxation energy is part of the total energy for a given system, it becomes obvious that verification of converging the atomic displacements as a function of supercell size is one of the critical steps that need to be accounted for in defect calculations.

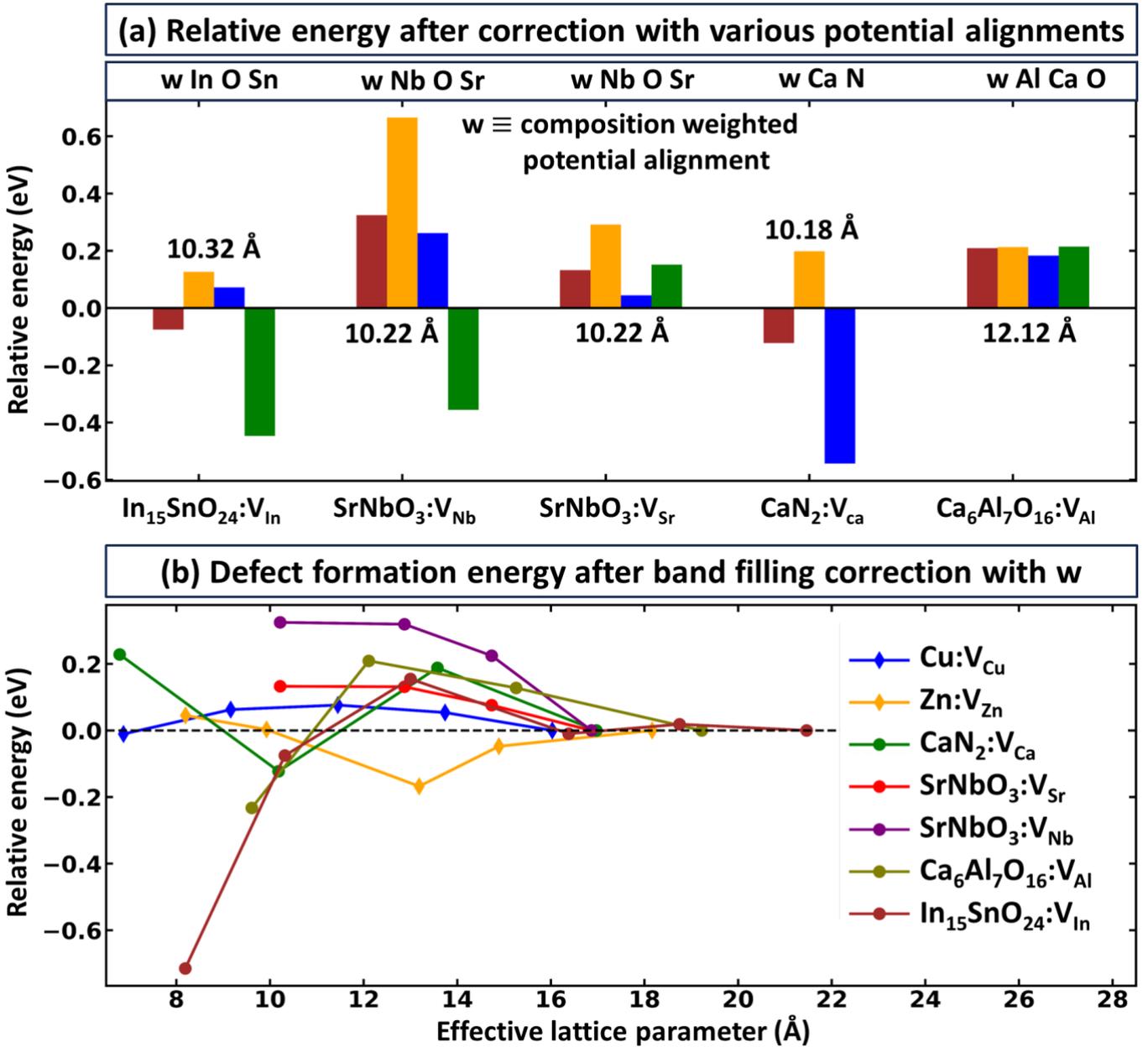

**Figure 6: Corrected defect formation energy for gapped metals.**
(a) Relative defect formation energy with respect to largest supercell after accounting for band-filling correction energy using various methods for alignment of electrostatic potentials (average of 3 most remote atoms of each species and the composition weighed (w) potential alignment). (b) Defect formation energy after band filling correction with w alignment as a function of the size of the system. Relative energies in a and b are calculated with respect to the defect formation energy of the largest supercell after band-filling correction.

**Conclusions:** In summary, using first-principles calculations, we provide a detailed analysis of defect formation in different types of metals. We show that not all metals are similar to each other and demonstrate the unique defect physics of gapped metals (i.e., compounds having internal band gaps similar to insulators with Fermi level within one of the main band edges). For instance, while the formation energy of point defects in Cu and Zn (i.e., conventional metals) can be routinely calculated using relatively small-size supercells, gapped metals (i.e., quantum materials exemplified herein by $Ca_6Al_7O_{16}$, $SrNbO_3$, $In_{15}SnO_{24}$, and

CaN₂) exhibit strong supercell size dependence of defect formation. This behavior is driven by the effect of point defects on their electronic structure. Thus, the pristine n-type gapped metal has the Fermi level inside the conduction band and a large internal band gap below it. Hence, the formation of an acceptor defect can result in a decay of a fraction of conducting electrons to the acceptor defect level (different relative faction for different supercells), resulting in the reduction of free carrier concentration and shifting the Fermi level. While in the concentrated limit, such behavior reproduces the real system, to describe the defect formation energy in gapped metal in the dilute limit, one should explicitly account for the change of band filling and its effect on defect energetics, which indeed can be done in the form of the post-process correction using potential alignment between defective and pristine systems made via utilizing composition-weighted alignment of electrostatic potential for most remote atoms. We emphasize as well that while including band-filling correction is needed for describing the physics of gapped metals, other factors, such as converging internal relaxation, may also have a significant effect on accurate calculations of defect formation energy. In this way, we provide a fundamental understanding of defects in gapped metals and pave the way for future research and their potential applications.

**Methods:** All calculations were performed using the Vienna Ab initio Simulation Package (VASP)[53-56], employing the Perdew-Burke-Ernzerhof (PBE)[57] functional. For plane wave basis, the cutoff energy levels were set to 550 eV for volume relaxation, and for structural relaxation, default values were used. The atomic relaxations were carried out (unless otherwise specified) until the intrinsic forces were below 0.01 eV/Å. The Γ-centered Monkhorst–Pack k-grid with a density of 10,000 per reciprocal atom was used for all main calculations (minimum k-points 2×2×2). The results were analyzed using pymatgen[58] and Vesta[59]. For regular metals, we used Cu (SG: 225)[60] and Zn (SG: 194)[61] as representative examples. While for the gapped metals, the calculations were performed for $SrNbO_3$ (SG: 62)[62], $Ca_6Al_7O_{16}$ (SG: 220)[63], $CaN_2$ (SG: 139)[64], and $In_{15}SnO_{24}$ (SG: 148)[65].

**Acknowledgment:** The authors thank the "ENSEMBLE3 - Centre of Excellence for nanophotonics, advanced materials and novel crystal growth-based technologies" project (GA No. MAB/2020/14) carried out within the International Research Agendas programme of the Foundation for Polish Science co-financed by the European Union under the European Regional Development Fund and the European Union's Horizon 2020 research and innovation programme Teaming for Excellence (GA. No. 857543) for support of this work. We gratefully acknowledge Poland's high-performance computing infrastructure PLGrid (HPC Centers: ACK Cyfronet AGH) for providing computer facilities and support within computational grant no. PLG/2023/016228 and for awarding this project access to the LUMI supercomputer, owned by the EuroHPC Joint Undertaking, hosted by CSC (Finland) and the LUMI consortium through grant no. PLL/2023/4/016319.

**AUTHOR DECLARATIONS**
**Conflict of Interest**
The authors have no conflicts to disclose.

**DATA AVAILABILITY**
The data that support the findings of this study are available from the corresponding author upon reasonable request.